\documentclass[aps,prb,twocolumn,floats]{revtex4}

\usepackage{ifthen}
\usepackage{ifpdf}

\usepackage{latexsym}
\usepackage{amssymb}
\usepackage{bm}

\ifpdf
\usepackage{graphicx}
\usepackage{epstopdf}
\else
\usepackage{graphicx}
\usepackage{epsfig}
\fi

\newcommand{\im}{\mbox{Im}}

\newcommand{\eexp}{\mbox{e}^}

\newcommand{\mass}{\mathsf{m}}

\newcommand{\tbox}[1]{\mbox{\tiny #1}}

\newcommand{\be}[1]{\begin{eqnarray}\ifthenelse{#1=-1}{\nonumber}{\ifthenelse{#1=0}{}{\label{e#1}}}}
\newcommand{\ee}{\end{eqnarray}} 

\newcommand{\hide}[1]{}

\newcommand{\Cn}[1]{\begin{center} #1 \end{center}}

\begin{document} 

\title{Quantum Stirring in low dimensional devices} 

\author{Itamar Sela and Doron Cohen}

\affiliation{
Department of Physics, Ben-Gurion University, Beer-Sheva 84005, Israel}


\begin{abstract}
A circulating current can be induced 
in the Fermi sea by displacing a scatterer, 
or more generally by integrating
a quantum pump into a closed circuit. 
The induced current may have either the 
same or the opposite sense with respect 
to the ``pushing" direction of the pump.  
We work out explicit expressions for the 
associated geometric conductance 
using the Kubo-Dirac monopoles picture, 
and illuminate the connection with the  
theory of adiabatic passage in multiple 
path geometry.
\end{abstract}

\maketitle


\section{Introduction}

The question ``how much charge is transported 
due to an adiabatic translation of a scatterer" 
has been raised in the context of an open geometry in Ref.\cite{avron}. 
The scatterer is a potential barrier whose 
location ${x=X_1}$ and transmission ${g_X=g(X_2)}$ 
at the Fermi energy are determined 
by gate controlled parameters~$X_1$ and~$X_2$.   
For the single mode ``wire" of Fig.~1a, using the 
Buttiker-Thomas-Pretre (BPT) formalism \cite{bpt, BPT2},  
one obtains the following result: 
\be{1}
Q= \left(1-g_{\tbox{X}}\right)\frac{e}
{\pi}k_{\tbox{F}} \ \Delta X_1
\ee
where $k_{\tbox{F}}=(2\mass E_{\tbox{F}})^{1/2}$ is the Fermi momentum, 
and $\Delta X_1$ is the translation distance of the scatterer. 
Ref.\cite{avron} has referred to this transport mechanism as ``snow plow" 
and pointed out that it should be regarded 
as the prototype example for quantum pumping: 
a full pumping cycle (Fig.~2a) would consist 
of translating the scatterer to the right, 
shrinking its ``size", pulling it back 
to the left, and restoring its original size.

Quantum stirring \cite{pml,pmt} is the operation 
of inducing a DC circulating current 
by means of AC periodic driving. 
This is naturally achieved by integrating  
a quantum pump in a closed circuit \cite{pms,pMB,JJ}.     
In particular Refs.\cite{pml,pmt}
have considered the same adiabatic ``snow plow" mechanism 
as described above and obtained for the 
model system of Fig.~1b the following result:  
\be{2}
Q= \left[\frac{(1-g_{\tbox{X}})g_{\tbox{V}}}
{g_{\tbox{X}}+g_{\tbox{V}}-2g_{\tbox{X}}g_{\tbox{V}}}\right]\frac{e}
{\pi}k_{\tbox{F}} \ \Delta X_1
\ee
where $g_{\tbox{V}}$ is the transmission of 
the ring segment that does 
not include the moving scatterer, 
as defined by its Landauer conductance 
if it were connected to reservoirs.

Eq.(\ref{e2}) is ``classical" in the Boltzmann sense   
because in its derivation the interference within 
the ring is ignored. The purpose of the present study 
is to derive quantum results for the stirring 
in a low dimensional device, 
where quantum mechanics has the most dramatic consequences.   
In particular we would like to illuminate the possibility 
of having a counter-stirring effect: 
by ``pushing" the particles (say) anticlockwise,  
one can induce a circulating current 
in the counter-intuitive (clockwise) direction.

\section{Outline}

As a preliminary stage we provide a simple pedagogical 
explanation of the counter-stirring effect by regarding 
the ``pushing stage" of the pumping cycle 
as an adiabatic passage in multiple path geometry \cite{cnb}.

For the actual analysis in the general case 
we use the Kubo-Dirac monopoles picture 
of \cite{pmc}. Within this framework the pumped charge $Q$ 
is determined by the flux of a $\bm{B}(X)$ field which is 
identified as the Berry-Kubo curvature \cite{berry1,avron2,berry2}.
We study both analytically and numerically 
how  this field looks like.  The results are illustrated  
in Figs.~2-4.  Summing the contributions of all the occupied levels
we get expressions for the geometric conductance~$G$.   
Integrating over a full pumping cycle we get results for~$Q$.

We derive practical estimates for the stirring  
which is induced due to the translation of 
either small ($g_{\tbox{X}}\sim1$) or a large ($g_{\tbox{X}} \ll 1$) scatterer, 
including the possibility of having $g_{\tbox{X}} \sim g_{\tbox{V}}$.
The dependence of~$Q$ on the size of the scatterer
is plotted in Fig.~5, where it is contrasted 
with the classical expectation, 
and compared with the analytical approximations.  
In the Summary we refer to the experimental measurement aspect.

\section{The counter-stirring effect}

The essence of the counter-stirring effect can be understood 
without the Kubo-Dirac monopoles picture by adopting  
the ``splitting ratio" concept of Ref.\cite{cnb}. 
Referring to Fig.~1b the translation 
of the scatterer to the right is effectively like lowering 
the potential floor in the left bond, and raising 
the potential floor in the right bond. This induces 
an adiabatic passage of a particle from the right to the left.
The particle has two possible ways to make 
the passage: either via the ``V" barrier (coupling $W_{12}^{\tbox{V}}$)  
or via the ``X" barrier (coupling $W_{12}^{\tbox{X}}$). 
The splitting ratio determines the fraction 
of the current that goes via the ``V" barrier:
\be{3}
\lambda(X_2) = \frac{W_{12}^{V}}{W_{12}^V+W_{12}^X} 
= {\frac{\sqrt{g_{\tbox{V}}}}{\sqrt{g_{\tbox{V}}}\pm\sqrt{g_{\tbox{X}}}}}
\ee
where the last equality is based on the analysis in Ref.\cite{pms}.
If $W_{12}$ were the classical rate of the transition,  
we would have ${0<\lambda<1}$ and the current would flow 
in accordance with our classical intuition. 
But $W_{12}^{\tbox{V}}$ and $W_{12}^{\tbox{X}}$ are {\em real} amplitudes 
that might have opposite signs if an odd level crosses 
an even level. Consequently if $|W_{12}^{\tbox{X}}|> |W_{12}^{\tbox{V}}|$ 
we get ${\lambda<0}$ which implies that a circulating 
current is induced in the counter-intuitive (clockwise) direction.
This does not come in any contradiction with the 
observation that the net transport (summing over both barriers) 
is still from right to left.

\section{The model Hamiltonian} 

Our model is a 1D coherent ring with a
fixed scatterer and a controlled scatterer.
The fixed scatterer is some potential barrier $V(x)$, 
and the controlled scatterer is modeled as a delta function 
whose position and transmission are determined  
by the control parameters $X_1$ and $X_2$ respectively.
The one particle Hamiltonian is:
\be{4}
{\cal H} = \frac{1}{2\mass}\hat{p}^2+
V(\hat{x})+X_2(t)\delta(\hat{x}-X_1(t))
\ee
with periodic boundary conditions 
over $x\in[-L/2,L/2]$ so as to have a ring geometry.
Below we further assume that both bonds are 
of similar length (${L{-}X_1 \sim X_1 \sim L/2}$). 
The current is measured through 
a section $x{=}x_0{=}+0$ at the fixed barrier,
and accordingly:
\be{5}
{\cal I} = \frac{e}{2\mass}\left(
\hat{p}\delta(\hat{x}-x_0)+
\delta(\hat{x}-x_0)\hat{p}\right)
\ee

We also define generalized forces
which are associated with the control parameters:
\be{0}
{\cal F}^1 =& -(\partial{\cal H}/\partial X_1)  &= X_2\delta'(\hat{x}-X_1)\\
{\cal F}^2 =& -(\partial{\cal H}/\partial X_2)  &=-\delta(\hat{x}-X_1)
\ee
For practical use it is more convenient to describe
the fixed scatterer by its scattering matrix, which can be written as:
\be{7}
{\bm S_{\tbox{V}}} &=&
\eexp{i\gamma_{\tbox{V}}}\left(
\begin{array}{cc}
-i\sqrt{1-g_{\tbox{V}}}\eexp{i\alpha_{\tbox{V}}} & \sqrt{g_{\tbox{V}}}\\
\sqrt{g_{\tbox{V}}} & -i\sqrt{1-g_{\tbox{V}}}\eexp{-i\alpha_{\tbox{V}}}
\end{array}
\right)
\ee
We study the case when
the model parameters are such that
the transmission of the fixed scatterer
is small ${(g_{\tbox{V}}\ll1)}$, 
the Fermi momentum is large (${k_{\tbox{F}}L \gg 1}$), 
and the controlled scatterer is translated
a distance $\Delta X_1$ that equals 
several Fermi wavelengths.

\section{The Kubo-Dirac picture} 

If we were changing the flux $X_3\equiv\Phi$ 
through the ring, the induced current would be given by 
Ohm law ${I=\langle\mathcal{I}\rangle=-G^{33}\dot{X_3}}$, 
where $G^{33}$ is the Ohmic conductance 
and $-\dot{X_3}$ is the electro-motive-force. 
Similarly for a variation of the parameter~$X_1$, 
the current is $I=-G^{31}\dot{X}_1$,
where $G^{31}$ is called the geometric conductance.
For two parameters driving one can write:
\be{8}
Q = \int I\mbox{d}t 
=-\oint_{\mbox{cycle}}\bm{G}\cdot\mbox{d}\bm{X}
= \oint\bm{B}\cdot\mbox{d}\bm{s}
\ee
where $\bm{G}=(G^{31},G^{32})$, 
and $\bm{X}=(X_1,X_2)$ 
and $\mbox{d}\bm{s}=(\mbox{d}X_2,-\mbox{d}X_1)$.
For a particle that evolves adiabatically  
in the level~$n$ we have ${G^{31}=B_2}$ and ${G^{32}=-B_1}$ where:
\be{11}
B_j^{(n)} =
\sum_{m(\neq n)}\frac{2 \
\im[{\cal I}^{}_{nm}]{\cal F}^{j}_{mn}}{(E_m-E_n)^2}
\ee
In fact $(B_1,B_2)$ are elements of 
the Kubo-Berry curvature \cite{berry1,avron2,berry2}
which one can regard as a fictitious 
magnetic field $\vec{B}=(B_1,B_2,B_3)$ 
in an embedding space ${X}=(X_1,X_2,X_3)$.
From the requirement of having well defined Berry phase 
it follows that the sources of $\vec{B}(X)$,  
which are located at points of degeneracy, 
are quantized, so called ``Dirac monopoles".

\section{The $X$ space} 

Due to the gauge symmetry ${\Phi\mapsto \Phi+2\pi\hbar/e}$ 
the Dirac monopoles are arranged as vertical chains (see Fig.~2(f)) 
[${\hbar{=}e{=}1}$]. 
Due to the time reversal invariance of our $\mathcal{H}(X_1,X_2)$ 
it follows that a Dirac chain is either a duplication of 
in-plane monopole at ${X_3=0}$ or off-plane monopole at ${X_3=\pi}$.
Let us find an explicit formula for the $(X_1,X_2)$ locations  
of these vertical chains.
The equation for the adiabatic energies  $E_n(X)$ 
is of the form $\cos(kL + \gamma)=\sqrt{g}\cos(\Phi)$ 
where~$g$ is the total transmission of the ring 
and~$\gamma$ is the total phase shifts of the scatterers 
(the fixed scatterers plus the moving scatterer). 
If $X_2$ is such that $g_{\tbox{X}}(E;X_2)=g_{\tbox{V}}(E)$ we can always 
find $X_1$ such that the total transmission would be ${g=1}$,  
which is the necessary condition for having a degeneracy.
Together with the equation ${k_{\tbox{E}} L + \gamma(E) = r\pi}$ 
with $r{=}\mbox{\small integer}$ 
this defines a set of energies ${E^{r}=(k^r)^2/2\mass}$ 
and associated values $X_2^{r}$ 
for which the $n{=}r$ level has degeneracy 
with the $n{=}r{+}1$ level provided $X_1$ is adjusted. 
To be more precise ${r{=}\mbox{\small even}}$  
are in-plane ($\Phi{=}0$) degeneracies, 
while ${r{=}\mbox{\small odd}}$ are 
off-plane ($\Phi{=}\pi$) degeneracies. 
The $X_1$ locations of these degeneracies 
are half De-Broglie wavelength apart (see Fig.3):
\be{12}
X_1^{r} = 
\frac{\alpha_{\tbox{V}}}{2k^r}
+\frac{L}{2}+
\left(\left[\frac{1}{2}\right]+\mbox{\small integer}\right) \ \frac{\pi}{k^r}
\ee
where the $[1/2]$ shift applies to in-plane degeneracies.     
The arrangement of the degeneracies in $X$ space 
is illustrated in Fig.2. For each $(X_1^{r},X_2^{r})$ 
we  have a vertical Dirac chain whose monopoles 
are formally like sources for the $\bm{B}$ field.

\section{Fermi occupation} 

If we have many body system 
of $N=\sum_n f_n$ particles,  
then $\bm{B} = \sum_n f_n\bm{B}^{(n)}$. 
At finite temperature each occupied level 
(except $n{=}1$) contributes {\em two} sets 
of $(X_1^{r},X_2^{r})$ chains, 
namely $r{=}n$ and $r{=}n{-}1$, 
which are associated with the $E_{n}=E_{n{\pm}1}$ degeneracies. 
By inspection of Eq.(\ref{e11}), taking into account 
that $\mathcal{I}_{nm}$ is antisymmetric, 
one observes that the net contribution of 
the $r$th set of Dirac chains is ${f_{r}-f_{r{+}1}}$.  
In particular for zero temperature Fermi occupation, 
the net contributions comes from only {\em one} set 
of chains which is associated with the degeneracies 
of the last occupied level with the first non-occupied 
level (Fig.~2bcd).

\section{Classical limit} 

At finite temperatures we can define 
the smeared probability distribution 
of the Dirac monopoles with respect 
to~$X_2$ as follows:
\be{0}
f(X_2) \ = \ \overline{\sum_r [f_{r}-f_{r{+}1}] \, \delta(X_2-X_2^{r})}
\ee
Disregarding fluctuations Eq.(\ref{e8}) implies 
a monotonic dependence of $Q$ on $X_2$ in qualitative 
agreement with Eq.(\ref{e2}). 
If the expression in the square brackets
of Eq.(\ref{e2}) were equal ${\int_0^{X_2} f(X')dX'}$, 
it would imply a quantitative agreement as well. 
In order to have this quantitative agreement 
we have to further assume that the distribution $f(X_2)$  
is determined by some chaotic dynamics in the 
scattering region which would imply erratic dependence 
of the $S$~matrix on the energy $E$. 
See \cite{pml} for further discussion.

\section{Quantum limit} 

Our interest below is in the opposite limit 
of zero temperature where $f(X_2)$ becomes 
a step function. Obviously in this limit 
a step like behavior of $Q$ versus $X_2$ would be 
a crude approximation.  
By inspection of Eq.(\ref{e11}) it follows   
that the result for ${G \equiv G^{31} = B_2}$ 
is very well approximated by 
\be{14}
G(X_1,X_2) = \frac{2 \ \im[{\cal I}^{}_{n,n{+}1}]{\cal F}^{1}_{n{+}1,n}}{(E_{n{+}1}-E_n)^2}
\ee
where $n$ is the last occupied level. 
This observation, as well as the associated  
analytical results which are based on it,  
have been verified against the exact numerical 
results of Figs.~3-4.
Below we derive explicit expressions for $G$ vs $X_1$ 
for both small and large values of $X_2$.
Our results for $Q$ are plotted in Fig.~5. 
Note that the dependence on $X_1$ has $\pi/k_{\tbox{F}}$ 
periodicity due to the $X$ space arrangement 
of the monopoles, and accordingly the integration 
gives~$Q \propto e k_{\tbox{F}} \Delta X_1 /\pi$, 
with a prefactor that we would like to estimate.

\section{Matrix elements}

The matrix elements of the current operator ${\cal I}$ 
and of the generalized force ${\cal F}^1$ are
\be{0}
{\cal I}_{nm} &=& i\frac{e}{2\mass}\left(
\partial\psi^{(n)} \ \psi^{(m)}-\psi^{(n)} \ \partial\psi^{(m)}\right)\\
{\cal F}^1_{mn} &=& 
-X_2\left(\overline{\partial\psi^{(m)}} \ \psi^{(n)}+
\overline{\partial\psi^{(n)}} \ \psi^{(m)}\right)
\nonumber\\
&=&
-\frac{1}{2\mass}
\left[\partial\psi_R^{(m)}\partial\psi_R^{(n)}-
\partial\psi_L^{(m)}\partial\psi_L^{(n)}
\right]
\ee
where $\overline{\partial\psi}=1/2(\partial\psi_L+\partial\psi_R)$ 
is the average derivative on the left and right
sides of the delta barrier, and the second
expression for ${\cal F}^1_{nm}$ was obtained
by using the matching conditions 
across the delta barrier.
The wave function 
is written as ${\psi(x) = C\sin(\varphi+kx)}$. 
We found that a very good approximation 
for ${\cal I}_{nm}$ with $m{=}n{+}1$ is    
\be{19}
{\cal I}_{nm} &=& \pm ie\frac{v_{\tbox{F}}}{L}\sqrt{g_{\tbox{V}}}
\ee
where the $+$ $(-)$ sign is for $n=$even (odd), 
and  $v_{\tbox{F}}=k_{\tbox{F}}/\mass$ is the velocity 
in the energy range of interest. 
For the calculation of ${\cal F}^1_{mn}$ 
and $E_{m}-E_{n}$ we have to distinguish 
between the two cases of small/large scatter.  
This means the small/large $X_2$ regimes 
where $g_{\tbox{X}} \sim 1$ or $\ll 1$.

\section{Translating a small scatterer} 

If the controlled scatterer is small, 
we treat it as a perturbation.  
For the energy splitting we get 
\be{0}
E_m{-}E_n \approx \frac{\pi}{L}v_{\tbox{F}}  
\mp\frac{2}{L}X_2 \cos(2k_{\tbox{F}}X_1)
\ee
where for notational convenience 
we take $X_1^{r}$ as the new origin.  
After some further algebra we get  
\be{0}
{\cal F}^1_{mn} = \pm X_2\frac{2k_{\tbox{F}}}{L} 
\cos\left(2k_{\tbox{F}}X_1\right){+}
X_2\frac{\pi}{L^2}
\ee
where the $\pm$ sign is as in Eq.(\ref{e19}).
The conductance can be written as 
\be{0}
G = \frac{e}{\pi}k_{\tbox{F}}\sum_{\nu=0}^\infty {\cal G}_\nu\cos\left(\nu \ 2k_{\tbox{F}}X_1\right)
\ee
where the coefficients of the leading non-negligible 
terms [the small parameter being $(1{-}g_{\tbox{X}})/g_{\tbox{X}}$] are
\be{190}
{\cal G}_0 &=& {\pm}2\sqrt{g_{\tbox{V}}}\left(
\frac{1}{k_{\tbox{F}}L}\sqrt{\frac{1-g_{\tbox{X}}}{g_{\tbox{X}}}}
{+} \frac{4}{\pi^2} \ {\frac{1-g_{\tbox{X}}}{g_{\tbox{X}}}}\right)\\
{\cal G}_1 &=& \frac{2}{\pi}\sqrt{g_{\tbox{V}}} \
\left(\frac{4}{k_{\tbox{F}} L}{\frac{1-g_{\tbox{X}}}{g_{\tbox{X}}}}{+}
2\sqrt{\frac{1-g_{\tbox{X}}}{g_{\tbox{X}}}}\right)
\ee
Upon integration we get
$Q=-e \, {\cal G}_0$
per half Fermi wavelength 
displacement of the scatterer.

\section{Translating a large scatterer}

If the controlled scatterer is large, most of the charge transfer
is induced during the avoided crossings (sharp peaks in Fig.~3 lower panel).
Consequently we use the two level approximation scheme 
of~\cite{pms} with $m=n{+}1$ leading to the results:
\be{0}
E_{m}-E_{n} &=& \frac{2}{L}v_{\tbox{F}} \ |{\bm R}| \\
{\cal F}^1_{mn} &=& \pm\frac{2}{L}\mass{v_{\tbox{F}}^2} \ \frac{R_2}{|{\bm R}|}
\ee
where the $\pm$ sign is as in Eq.(\ref{e19})
and the dimensionless distance in ${\bm X}$ space from
the degeneracy point is:
\be{0}
{\bm R} &=&\left(2k_{\tbox{F}}(X_1-X_1^{r}),\
\frac{\sqrt{g_{\tbox{V}}}}{\lambda(X_2)}\right)
\ee
Accordingly the conductance is
\be{30}
G = e\frac{k_{\tbox{F}}}{\sqrt{g_{\tbox{V}}}} \ \frac{R_2}{|{\bm R}|^3}
\ee
Integrating over $X_1$ we get $Q=e\lambda(X_2)$ 
per half Fermi wavelength displacement of the scatterer, 
as expected from the ``splitting ratio" argument.

\section{Summary}

The integration of a two-terminal quantum pump 
in a closed circuit is not a straightforward procedure. 
Due to interference the pumped charge~$Q$ 
would not be the same as in the Landauer/BPT setup, 
and even the sense of the induced current might 
be reversed. The most dramatic consequences 
would be observed in low dimensional devices. 
For this reason we have analyzed in this paper the prototype 
problem of ``pushing" a current by translating a scatterer 
in a single mode wire.  
We have obtained explicit results for the $\bm{B}$ field,  
which determines the geometric conductance~$G$, 
and consequently the~$Q$ of a closed pumping cycle. 
We also illuminated the counter-stirring effect using 
the splitting ratio concept of adiabatic passage 
in multiple path geometry.  
   
A few words are in order regarding the measurement procedure 
and the experimental relevance. It should be clear 
that to measure current in a closed circuit 
requires special techniques \cite{orsay,expr1,expr2}. 
These techniques are typically used in order 
to measure persistent currents, 
which are zero order (conservative) effect, 
while in the present paper we were discussing driven 
currents, which are a first-order (geometric) effect. 
It is of course also possible to measure the dissipative 
conductance (as in~\cite{orsay}).  
During the measurement the coupling to the system 
should be small. These are so called {\em weak measurement} 
conditions. More ambitious would be to measure 
the counting statistics, i.e. also the second moment 
of $Q$ as discussed in \cite{cnb,cnz}  which is 
completely analogous to the discussion of noise measurements 
in open systems \cite{levitov,nazarov}.
Finally it should be pointed out that the formalism 
above, and hence the results, might apply to 
experiments with superconducting circuits (see \cite{JJ}).


\acknowledgments

This research was supported by grants from 
the USA-Israel Binational Science Foundation (BSF), 
and from the Deutsch-Israelische Projektkooperation (DIP).




\clearpage

\begin{figure}[h]
\Cn{\includegraphics[width=1\hsize]{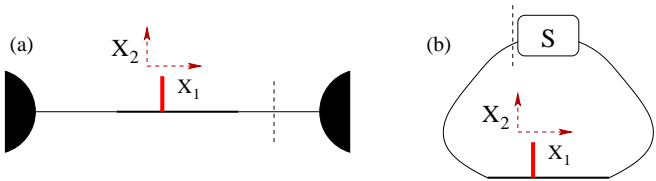}}
\caption{
In the case of an open geometry 
the pumping device is connected 
between two unbiased reservoirs (panel (a)), 
while in the present study 
it is integrated into a ring (panel (b)).
The induced current is measured 
through a section indicated by a dashed line.
See the text for further details.}
\end{figure}

\ \\ \ \\

\begin{figure}[h]
\Cn{\includegraphics[width=1.0\hsize]{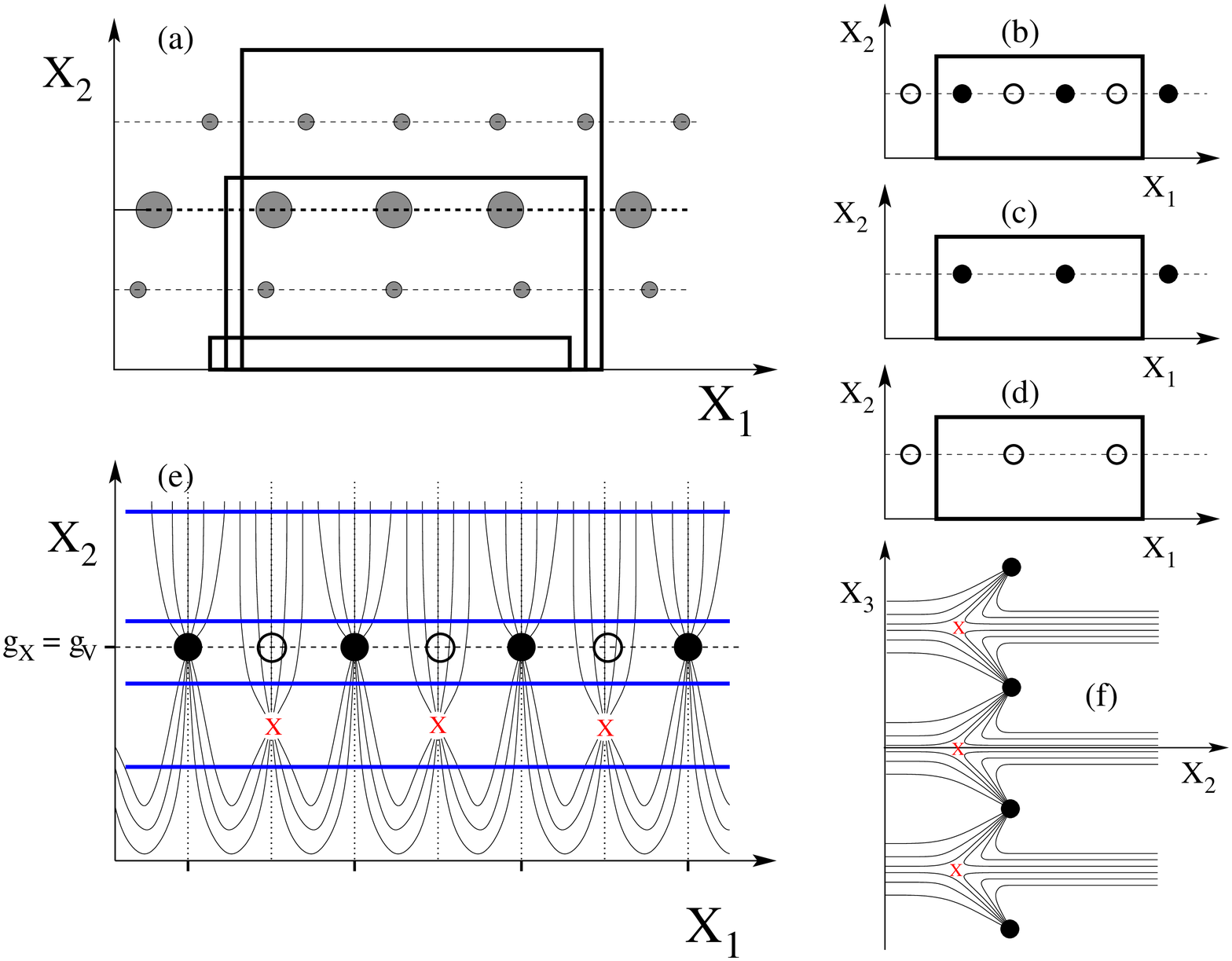}}
\caption{
(a) Three representative pumping cycles. 
Several sets of Dirac monopoles may have 
non-zero weight depending on the occupation. 
Panel (b) is for single level occupation 
where two sets have non zero weight, 
while either (c) or (d) are for zero temperature Fermi occupation. 
Filled (hollow) circles indicate in(off)-plane monopoles.
Panels (e-f) give a detailed illustration 
of the associated~$\mathbf{B}$ field, 
as implied by the numerical findings of Figs.3-4.
The $X_1$ tick marks in~(e) are half De-Broglie spaced, 
while the horizontal blue lines are paths
for which numerical results are presented in Figs.~3-4.}
\end{figure}

\begin{figure}[h]
\Cn{
\includegraphics[width=0.75\hsize]{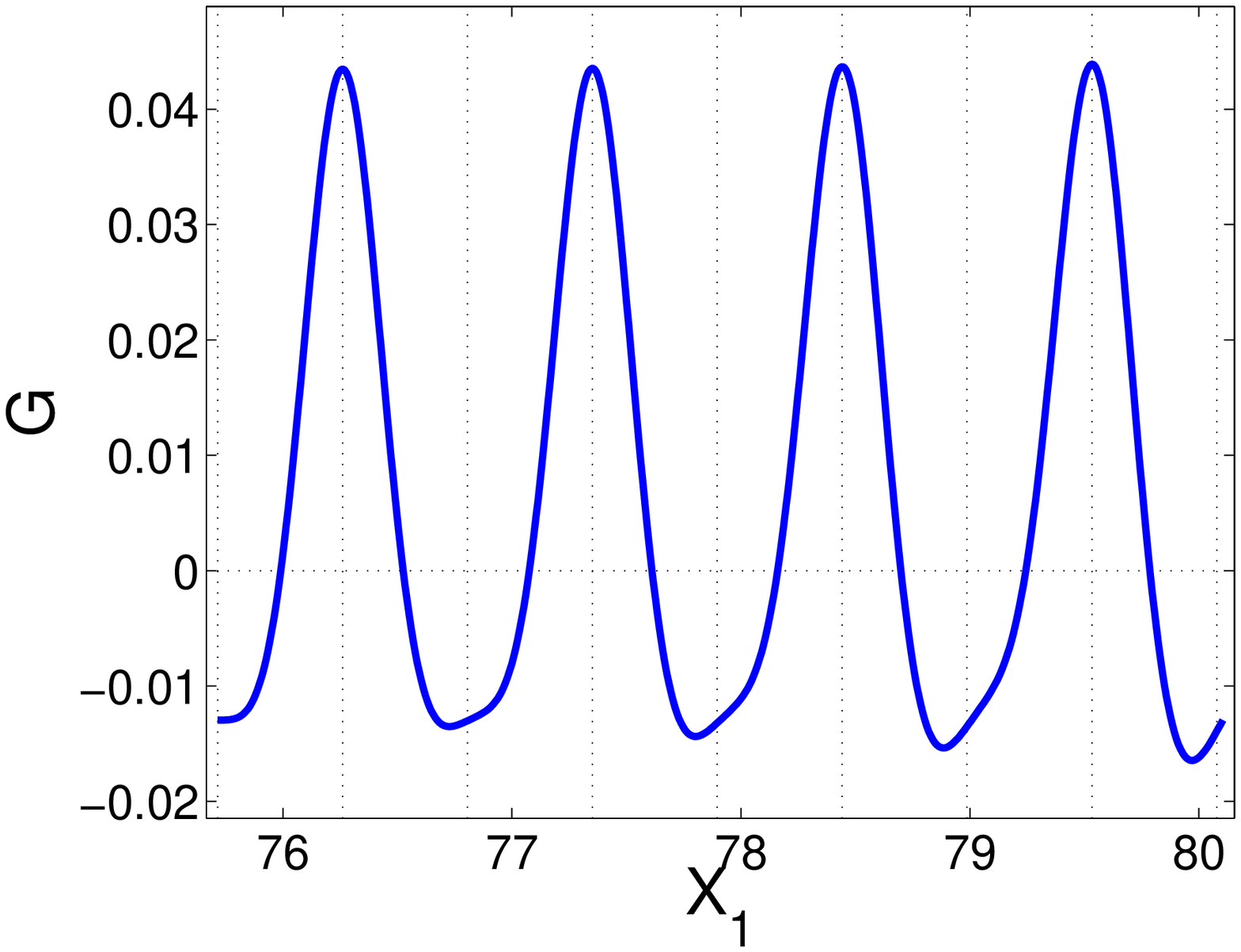} \\
\includegraphics[width=0.75\hsize]{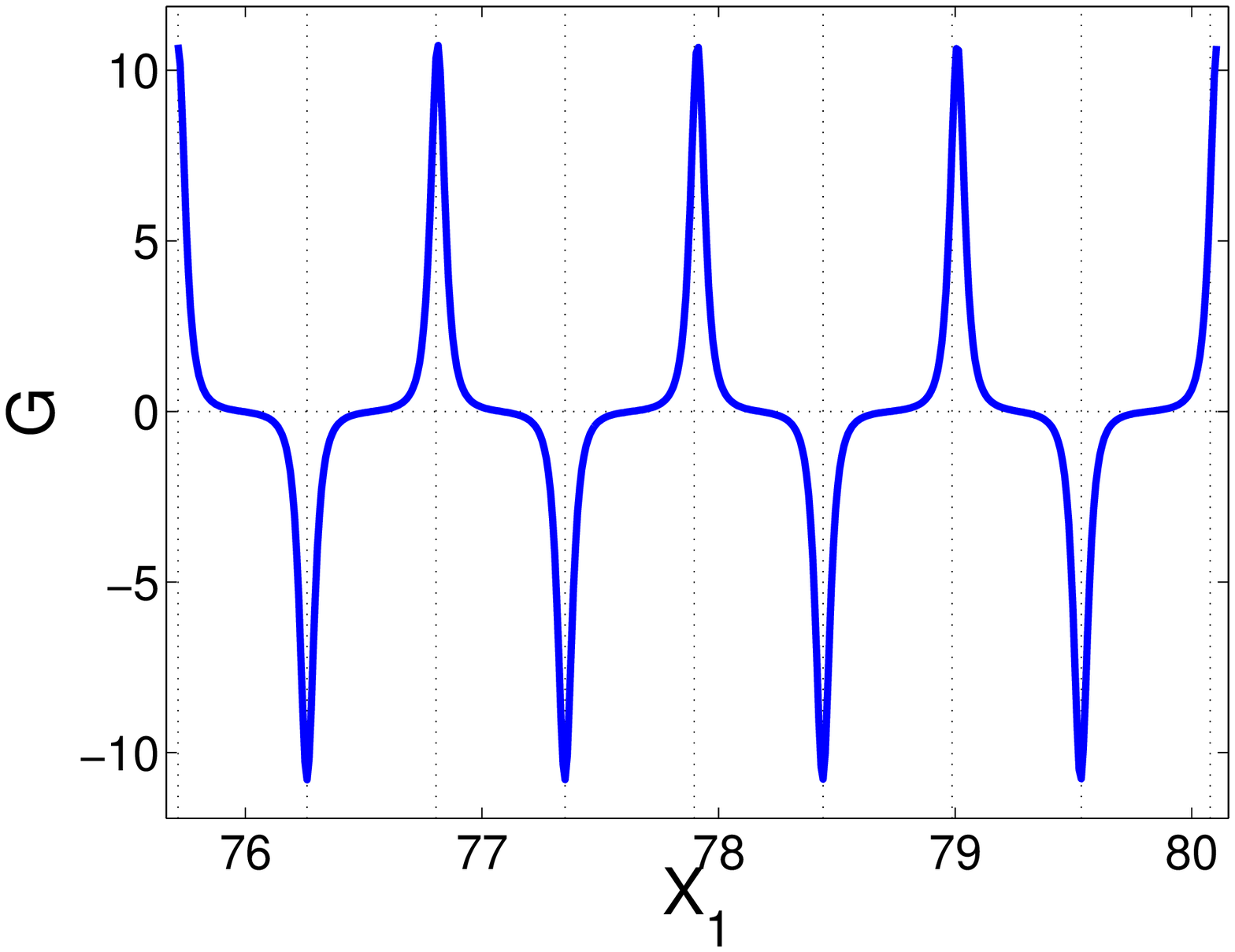}
}
\caption{
The conductance $G{=}G^{31}$ of Eq.\ref{e11}   
is numerically calculated for 
a ring of length $L{=}151.43$ 
with $V(x){=}U\delta(x)$ where $U{\sim}10$. 
We consider single level occupation $n{=}138$. 
At this energy ${g_{\tbox{V}}{=}0.06}$. 
The upper (lower) panel is for translation 
of a very small (large) scatterer 
with ${g_{\tbox{X}}{=}0.98}$ (${g_{\tbox{X}}{=}8\cdot 10^{-8}}$) 
corresponding to the lower (upper) horizontal blue paths 
that are indicated in Fig.~2e (same $X_1$ axis).}
\end{figure}

\begin{figure}[h]
\Cn{
\includegraphics[width=0.8\hsize]{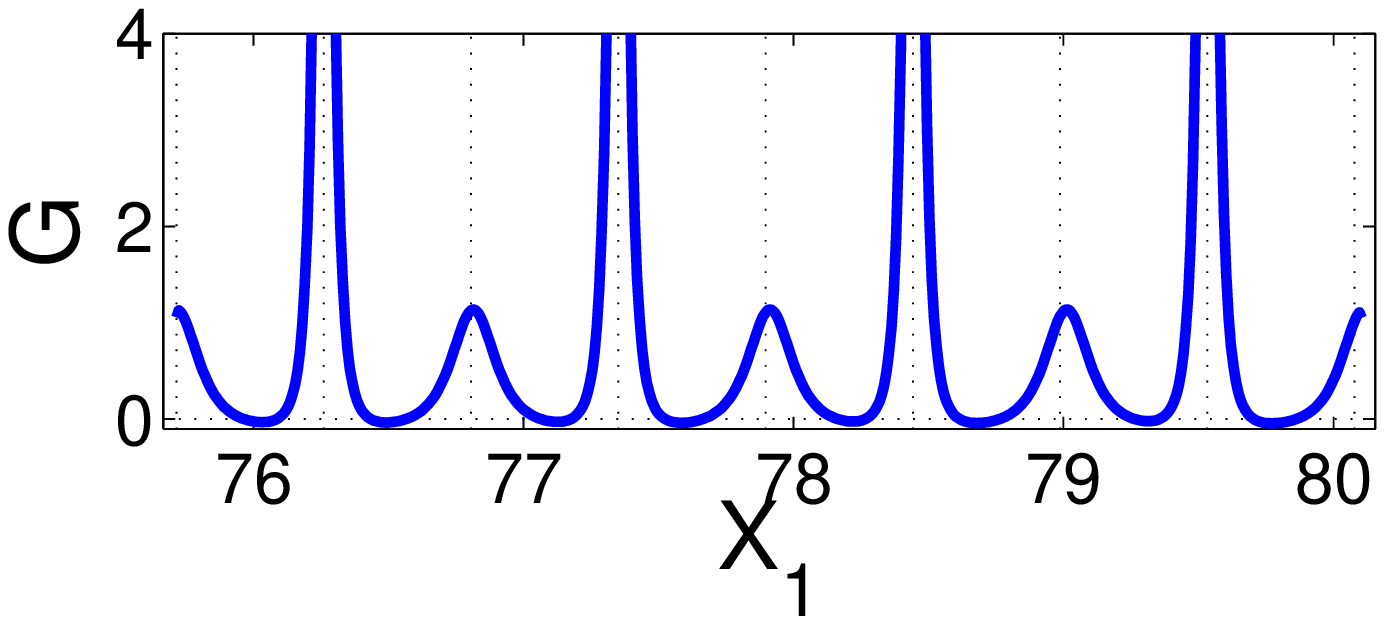} \\
\includegraphics[width=0.8\hsize]{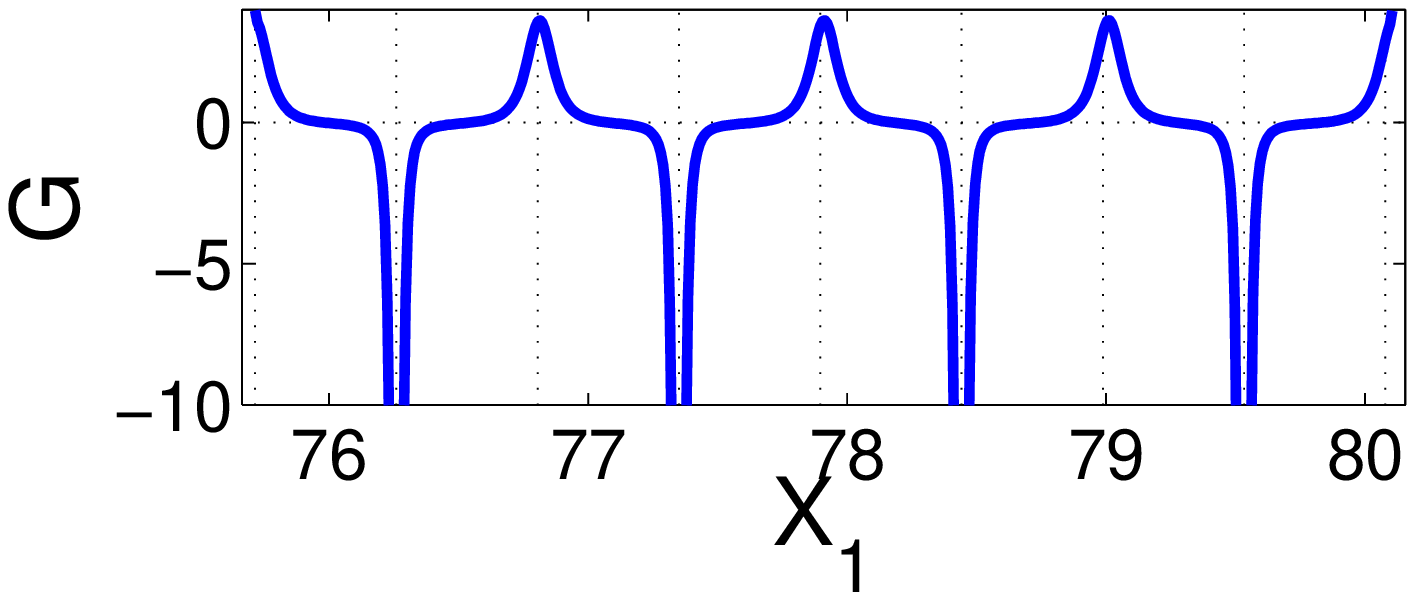}
}
\caption{
Additional plots of the conductance~$G$ 
as calculated in the previous figure. 
The upper (lower) panel is for translation 
of a scatterer with ${g_{\tbox{X}}{=}0.20}$ (${g_{\tbox{X}}{=}0.03}$)
corresponding to the horizontal blue paths 
in Fig.~2e that go below (above) 
the ${g_{\tbox{X}}{=}g_{\tbox{V}}}$ axis. 
Note that the large peaks are positive (negative)
while the small positive peaks switch sign only 
when the scatterer is lowered further. 
This indicates that the field lines bend in the~$X_3$ 
direction as illustrated in Fig.2f.}
\end{figure}

\clearpage

\begin{figure}[h]
\Cn{
\includegraphics[width=\hsize]{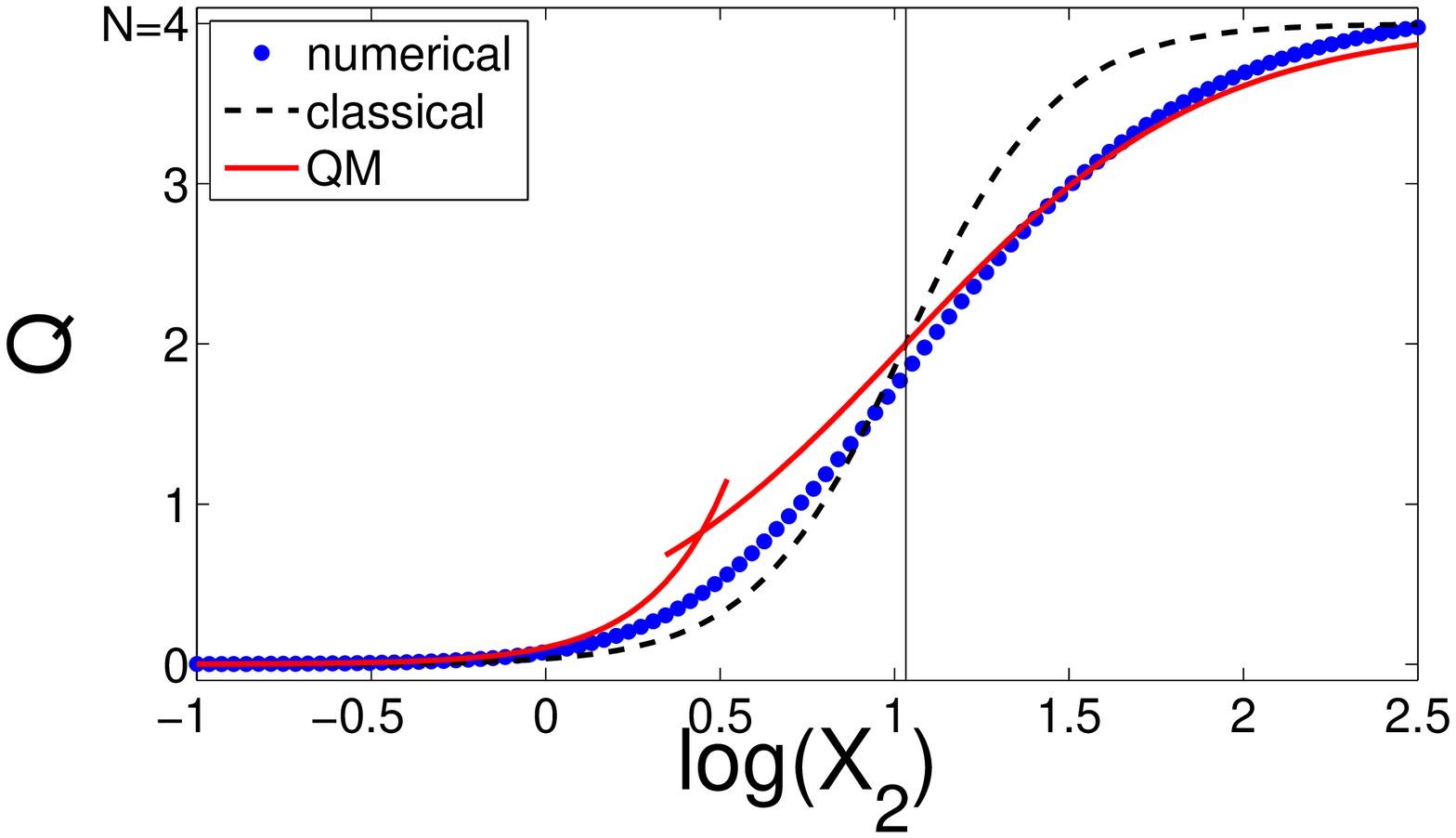}\\
\includegraphics[width=0.9\hsize]{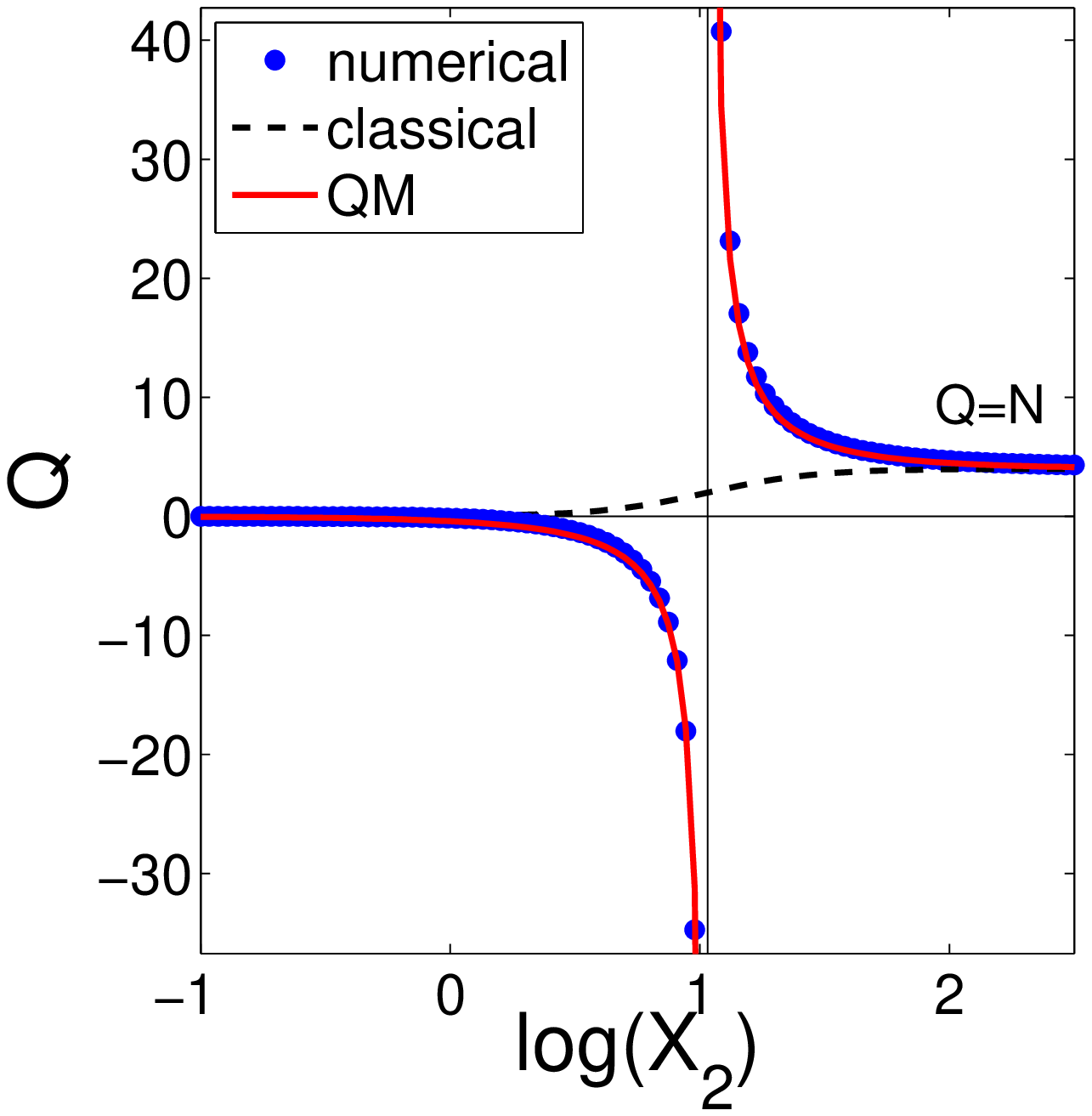}\\
\includegraphics[width=0.9\hsize]{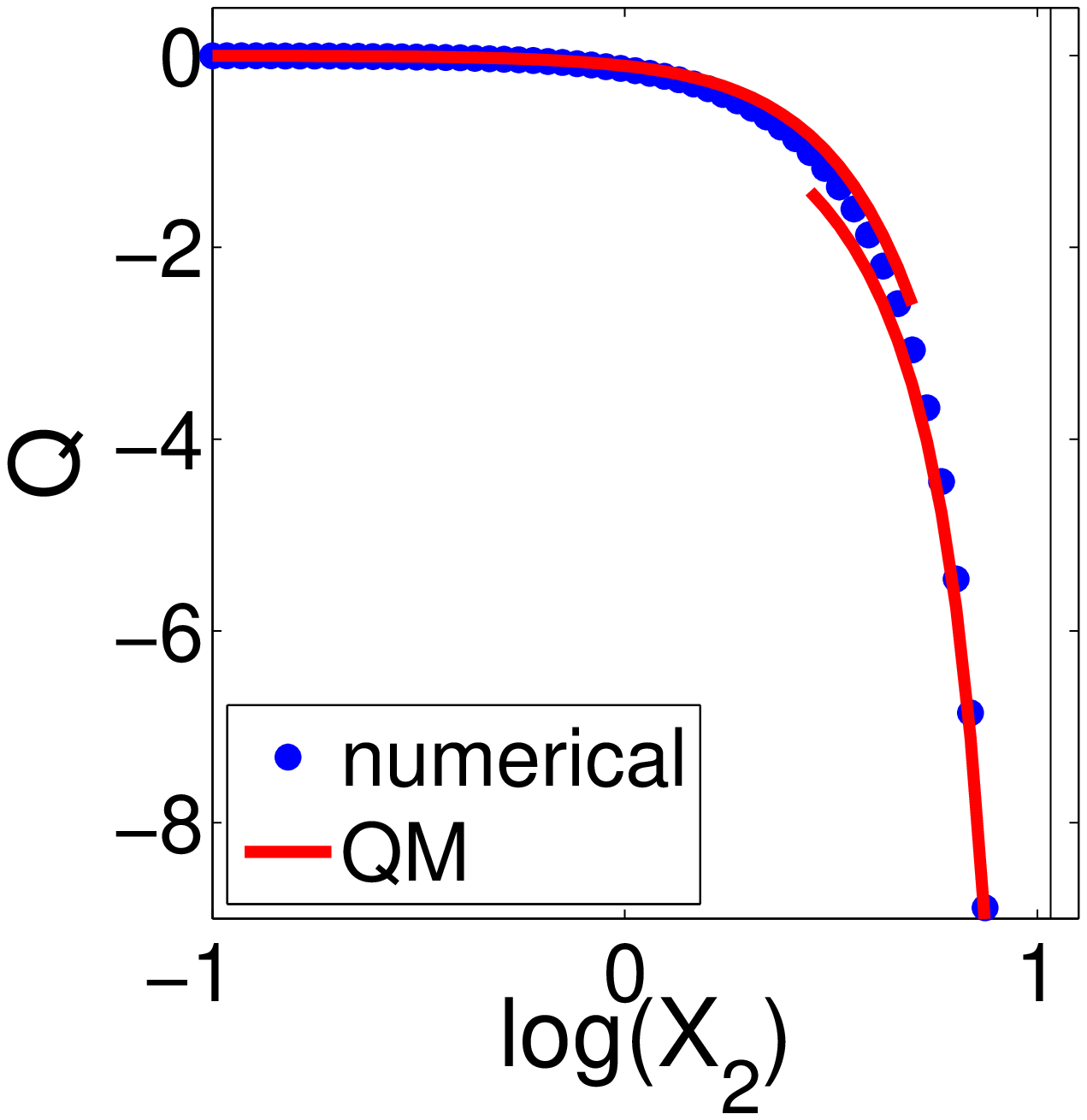}
}
\caption{
Numerical calculation of $Q$ for zero temperature Fermi occupation: 
all the levels are occupied up to $n{=}137$ (upper panel), 
and up to $n{=}138$ (lower regular and zoomed panels).
The model parameters are the same as in Fig.~3.
The integration was carried out along segments similar
to the paths shown in Fig.~2e where $g_{\tbox{X}}$
varies between $g_{\tbox{X}}{=}8.30\cdot10^{-5}$
and $g_{\tbox{X}}{=}0.45$.
For sake of comparison we display both the analytical 
classical (Eq.\ref{e2}) and quantum results (Eqs.\ref{e190}\&\ref{e30}).   
The value of $X_2$ for which $g_{\tbox{X}}{=}g_{\tbox{V}}$
is indicated by a vertical line.}
\end{figure}

\clearpage


\begin{thebibliography}{0}



\bibitem{avron} 
J. E. Avron, A. Elgart, G. M. Graf and L. Sadun,  Phys. Rev. B 62 (2000) 10618

\bibitem{bpt} 
M. Buttiker, H. Thomas and A. Pretre, Z. Phys. B 94 (1994) 133

\bibitem{BPT2}
P. W. Brouwer, Phys. Rev. B 58 (1998) 10135 

\bibitem{pml}
G.~Rosenberg and D.~Cohen, J. Phys. A 39 (2006) 2287 

\bibitem{pmt}
D. Cohen, T. Kottos and H. Schanz, Phys. Rev. E 71 (2005) 035202(R)

\bibitem{pms} 
I. Sela and D. Cohen, J. Phys. A 39 (2006) 3575 

\bibitem{pMB}
M.~Moskalets and M.~B{\"u}ttiker, 
Phys. Rev. B 68 (2003) 161311 

\bibitem{JJ}
M. Mottonen, J. P. Pekola, J. J. Vartiainen,
V. Brosco and F. W. J. Hekking, Phys. Rev. B 73 (2006) 214523 


\bibitem{cnb}
M. Chuchem and D. Cohen, J. Phys. A {\bf 41}, 075302 (2008).

\bibitem{pmc}
D. Cohen, Phys. Rev. B 68 (2003) 155303

\bibitem{berry1}
M.V. Berry, Proc. R. Soc. Lond. A 392 (1984) 45 

\bibitem{avron2}
J. E. Avron, A. Raveh and B. Zur, Rev. Mod. Phys. 60 (1988) 873

\bibitem{berry2}
M.V. Berry and J.M. Robbins, Proc. R. Soc. Lond. A 442 (1993) 659 




\bibitem{orsay} 
Measurements of currents in arrays 
of closed rings are described by:  \
B. Reulet M. Ramin, H. Bouchiat and D. Mailly, 
Phys. Rev. Lett. {\bf 75}, 124 (1995). 

\bibitem{expr1} 
Measurements of currents in individual closed rings 
using SQUID is described in: \
N.C. Koshnick, H. Bluhm, M.E. Huber, K.A. Moler, Science 318, 1440 (2007).

\bibitem{expr2} 
A new micromechanical cantilevers technique for measuring currents 
in closed rings is described in:  
A.C. Bleszynski-Jayich, W.E. Shanks, R. Ilic, J.G.E. Harris, arXiv:0710.5259.



\bibitem{cnz}
M. Chuchem and D. Cohen, Phys. Rev. A {\bf 77}, 012109 (2008).

\bibitem{levitov}
L.S. Levitov and G.B. Lesovik, JETP Letters {\bf 58}, 230 (1993).

\bibitem{nazarov}
Y.V. Nazarov and M. Kindermann, European Physical Journal B {\bf 35}, 413 (2003).


\end{thebibliography}
\end{document}